\documentclass[twocolumn,showpacs,prc]{revtex4}

\usepackage{graphicx}
\usepackage{dcolumn}
\usepackage{bm}

\begin{document}

\title{Sharp change over from compound nuclear fission to shape dependent quasi fission \\}

\author{T. K. Ghosh, K. Banerjee, C. Bhattacharya, S. Bhattacharya, S. Kundu, P. Mali, J. K. Meena, G. Mukherjee, S. Mukhopadhyay, and T. K. Rana. } 
\affiliation{Variable Energy Cyclotron Centre,  1/AF  Bidhan  Nagar,
  Kolkata  700 064, India}
\author{P. Bhattacharya}  \affiliation{Saha Institite of Nuclear Physics,  1/AF  Bidhan  Nagar,
  Kolkata  700 064, India.}
\author{K. S. Golda}  \affiliation{Inter University Accelerator Centre, New Delhi-110067,
 India.}

\date{\today}

\begin{abstract}
Fission fragment mass distribution has been measured from the decay of 
$^{246}$Bk nucleus populating via two entrance channels with slight difference 
in mass asymmetries but belonging on either side of the Businaro Gallone 
mass asymmetry parameter. Both the target nuclei were deformed. Near the Coulomb barrier, at similar excitation energies the width of the fission fragment mass distribution 
was found to be drastically different for the $^{14}$N + $^{232}$Th reaction compared 
to the $^{11}$B + $^{235}$U reaction. The entrance channel mass asymmetry was found 
to affect the fusion process sharply.

\end{abstract}


\maketitle

In recent years considerable interest has been observed in studying the fusion of two heavy ions to reach the island of stability in the super heavy nuclei \cite{Ognessian}.The experimental studies are not only aimed at actual production of the super-heavy nuclei, but also on the study of the 
fusion-fission process of the super heavy nuclei \cite{ItkisNPA}. The direct evidence of the fusion leading to a super heavy nucleus is the production of the evaporation residue, while in most of the cases the super heavy nucleus decays by spontaneous fission. Hence the study of the fusion-fission reactions becomes important to fix the initial conditions to maximize the probability for production as well as survival of the super heavy nucleus. 

It was generally believed that the fusion-fission process obeys statistical laws as the fission follows complete equilibration of the fused system of the colliding heavy ions in the entrance channel. Hence the fission reaction observables, such as the fragment mass and angular distributions, would follow statistical laws with no imprint of the entrance channel variables, such as the mass asymmetry, excitation energy or the overall fissility of the composite system. However, in several systems, remarkable departures from the statistical theory predictions in fission fragment angular distributions have been observed with evidence of predominance of  non-compound fission channels, particularly in near Coulomb barrier energies, with dependence on the entrance channel variables \cite{KailashPR}. This departure from theory was explained in terms of pre-equilibrium fission model \cite{RamPRL85} where it was argued that the composite system breaks before complete equilibration in K-degree (K being the projection of total angular momentum on the symmetry axis) of freedom.

Anomalous behaviour in fission fragment mass distributions was also observed \cite{myPLB,Renju08} around the Coulomb barrier. It has been suggested \cite{MollerNature} that in the multi-dimensional potential energy landscape, in the above barrier energies, the system preferentially moves over a  fusion  barrier to reach a composite system that reaches a meadow of  evaporation residues. After emission of  particles, the composite system undergoes binary fission following shape oscillations over a unconditional mass-symmetric fission barrier. However, as the excitation energy is lowered (i.e, at the near or below barrier energies), this path gets progressively blocked. But the system, in the case of the highly deformed entrance channel nucleus (either target or projectile or both), reaches a mass-asymmetric fission barrier following a ridge along the fusion-hill and produces events with  increased width of the mass distribution, which could be observed in experiments. We reported such observation for deformed target of  $^{232}$Th in reaction  with $^{19}$F, $^{16}$O and $^{12}$C \cite{myPLB}. The reaction mechanism as described above is termed as a nuclear orientation dependent quasi-fission reaction \cite{HindePRL95}, in which the projectiles in contact with the tip region of the prolate target give rise to the quasi-fission events. The probability of such reaction increases for mass flow preferentially from the heavier mass to the lighter mass to form the composite system. 

Evidence of quasi-fission for symmetric reaction populating low fissility composite system has been reported very recently \cite{Rafiei}. Since the super heavy nuclei are of higher fissility and to be produced at lower excitation energy, thus study of fusion-fission dynamics of a higher fissility composite system near the Coulomb barrier (i.e, at low excitation energy  with considerable fission cross section) is of prime interest. In this communication, we report a sharp dependence of the entrance channel mass asymmetry on the fusion-fission process in the production of the fusion of two heavy ions near the Coulomb barrier to produce  a highly fissile nucleus $^{246}$Bk. 

The heavy nucleus was produced with the reactions  $^{11}$B + $^{235}$U and $^{14}$N + $^{232}$Th . The beam energies were judiciously chosen to populate the $^{246}$Bk nucleus at similar excitation energies. The widths of the fragment mass distribution were measured. In an earlier measurement \cite{BeheraRapid} of fragment angular distribution for these two systems, deviations from statistical model predictions were observed only for the $^{14}$N + $^{232}$Th system, contrary to the claim of  a very recent dynamical cluster-decay model calculation \cite{Singh08}. The model calculation predicts that contribution from the non-compound fission channel should be higher for the  $^{11}$B + $^{235}$U compared to the reaction $^{14}$N + $^{232}$Th.

It is to be noted that both the $^{235}$U and $^{232}$Th targets chosen in our experiment to produce 
 $^{246}$Bk nucleus were deformed. However, the entrance channel mass asymmetries in two reactions were on either side of the Businaro Gallone mass asymmetry parameter value ($\alpha_{BG}=0.893$) \cite{Businaro}. It is of interest to note that in the entrance channel, for a nominal change in the entrance channel mass-asymmetry, the initial mass flow is from the lighter to heavier mass in the system  $^{11}$B + $^{235}$U ($\alpha=0.911$), while it is in the reverse direction, i.e., from the heavier to lighter mass in the $^{14}$N + $^{232}$Th ($\alpha=0.886$) system. Any inequality of the widths of the mass distribution in the decay of the $^{246}$Bk for the  systems with matched excitation and angular momentum  in the  entrance channel would be a confirmation of the effect of non-compound fusion-fission paths in the multi-dimensional potential energy surfaces. We have also used the reaction of  $^{14}$N + $^{197}$Au as a reference reaction, which is expected to be almost pure fusion-fission reaction at the excitation energies covered in the experiment.

We have in previous communications \cite{myRapid2,myPLB} established that measurement of the width of the mass distribution is a powerful tool to probe the reaction mechanism in fusion-fission  process. We employed a double armed time of flight spectrometer to simultaneously detect the complimentary fission fragments, rejected those events which showed incomplete fusion and determined the mass distribution for the purely fusion-fission process. The experimental arrangements and the data analysis procedure were described in detail in earlier reports \cite{myNIM}.  The experiments were carried out using pulsed $^{14}$N and $^{11}$B beam of width about 1.1 ns, with a pulse separation of 250 ns, from the 15UD Pelletron at the Inter University Accelerator Centre (IUAC), New Delhi. Targets of $^{197}$Au (self supporting), $^{232}$Th (on 200 {$\mu $}g/cm$^2$ Al backing) and $^{235}$U (on 300 {$\mu $}g/cm$^2$ Ni backing) of thickness 500 {$\mu $}g/cm$^2$  were used. Targets were placed at an angle of $45^{\circ}$ to the beam.  The fission fragments were well separated from elastic and quasi-elastic reaction channels, both from the event-time and energy loss spectra in the detectors. The fission fragments from complete fusion events followed by fission  were exclusively selected from the correlation  of the velocities of the fissioning  system in the beam direction ($V_{par}$) relative to the recoil  of the fused system  and the velocity perpendicular to the reaction plane  ($V_{perp}$), as well as the correlation of the polar  and azimuthal angles of the fragments ({$\theta$},{$\phi$) with respect to the beam axis. A typical measured distributions of the complementary fission events for the system $^{14}$N + $^{232}$Th at $E_{cm}$ = 77.3 MeV are shown in Fig.~\ref{fig:fig1}. To classify the events which solely come from fission following full momentum transfer with more than $99\%$ certainty, a cut was used as shown in the figure by (red) rectangle in the spectra of velocity distributions of the fissioning nucleus. The masses were determined from the angles, momentum and the recoil velocities for each event.

\begin{figure}
\includegraphics*[scale=0.3, angle=0]{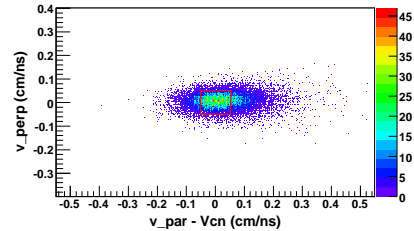}
\caption{\label{fig:fig1}~(Color online) Measured distributions of 
velocity of the fissioning nuclei formed in the reaction $^{14}$N+$^{232}$Th  at $E_{\rm c.m.} = 77.3$ MeV. The (red) rectangle  indicates the gate used to select 
the fusion-fission events for mass determination.} 
\end{figure}

\begin{figure}
\includegraphics*[scale=0.3]{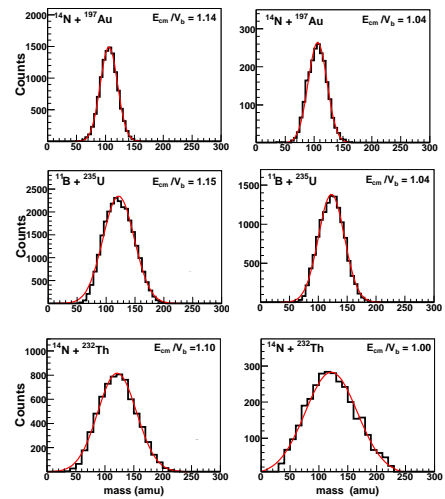}
\caption{\label{fig:fig2}~(Color online) Measured mass distributions for the reactions 
$^{14}$N+$^{197}$Au, $^{11}$B+$^{235}$U and $^{14}$N+$^{232}$Th near and above the 
Coulomb barrier. The Gaussian fits are shown by (red) solid lines.}
\end{figure}

Representative mass distributions, near and above the Coulomb barrier energies are shown in Fig.~\ref{fig:fig2} for $^{14}$N + $^{197}$Au,  $^{11}$B+$^{235}$U and  $^{14}$N + $^{232}$Th systems. It can be observed that measured mass distributions are well fitted with single Gaussian distributions at all energies.  The variation of the standard deviation ($\sigma_{m}$) of the fitted Gaussian to the experimental masses as a function of $E_{\rm c.m.}/V_b$ }, where the $E_{\rm c.m.}$ is the beam energy in center of mass system and $V_b$ is the Coulomb barrier, are shown in Fig 3 for all three measured systems. It is seen that for $^{14}$N+$^{197}$Au and $^{11}$B+$^{235}$U reactions  the variation of $\sigma_{m}$ is smooth across the Coulomb barrier. However a significant difference in the trend of the variation of $\sigma_m$ is observed for the $^{14}$N + $^{232}$Th system, it shows a sudden increasing trend around  the Coulomb barrier energies. Without delving into the reason for different behaviour for the  $^{14}$N + $^{232}$Th system, it was also noticed that for this system, events following the incomplete-fusion followed by fission is also significant compared to that for other two systems. Since admixture of transfer induced fission events, which increase with lowered beam energy, may give rise to a wider mass distribution, extra care was taken while analysing the data of $^{14}$N + $^{232}$Th system at energies where we observed a sudden  increase in the width  of the mass distributions as shown in Fig.~\ref{fig:fig3}. In Fig.~\ref{fig:fig4} we show the folding angle distribution  of  all the fragments for $^{14}$N + $^{232}$Th at $E_{cm}/V_{b}$= 1. The gate on the velocity correlations for fragments as shown in Fig.~\ref{fig:fig1} reduces the possible contribution of the transfer induced events to below $1\%$. To quantitatively evaluate the change in width of the mass distribution of even this possible small admixture of transfer fission events,  an additional conditionality was introduced through the imposition of a gate ($163^{\circ}$ to $174^{\circ}$)  around the peak  at $165.4^{\circ}$ which corresponds to that expected for the full momentum transfer symmetric fission events. This action changed the width of mass distribution which was smaller than the statistical error in $\sigma_m$.  This clearly establishes that the sudden increase in $\sigma_m$ is not due to admixture of  transfer induced events.

\begin{figure}
\includegraphics*[scale=0.3,angle=-90]{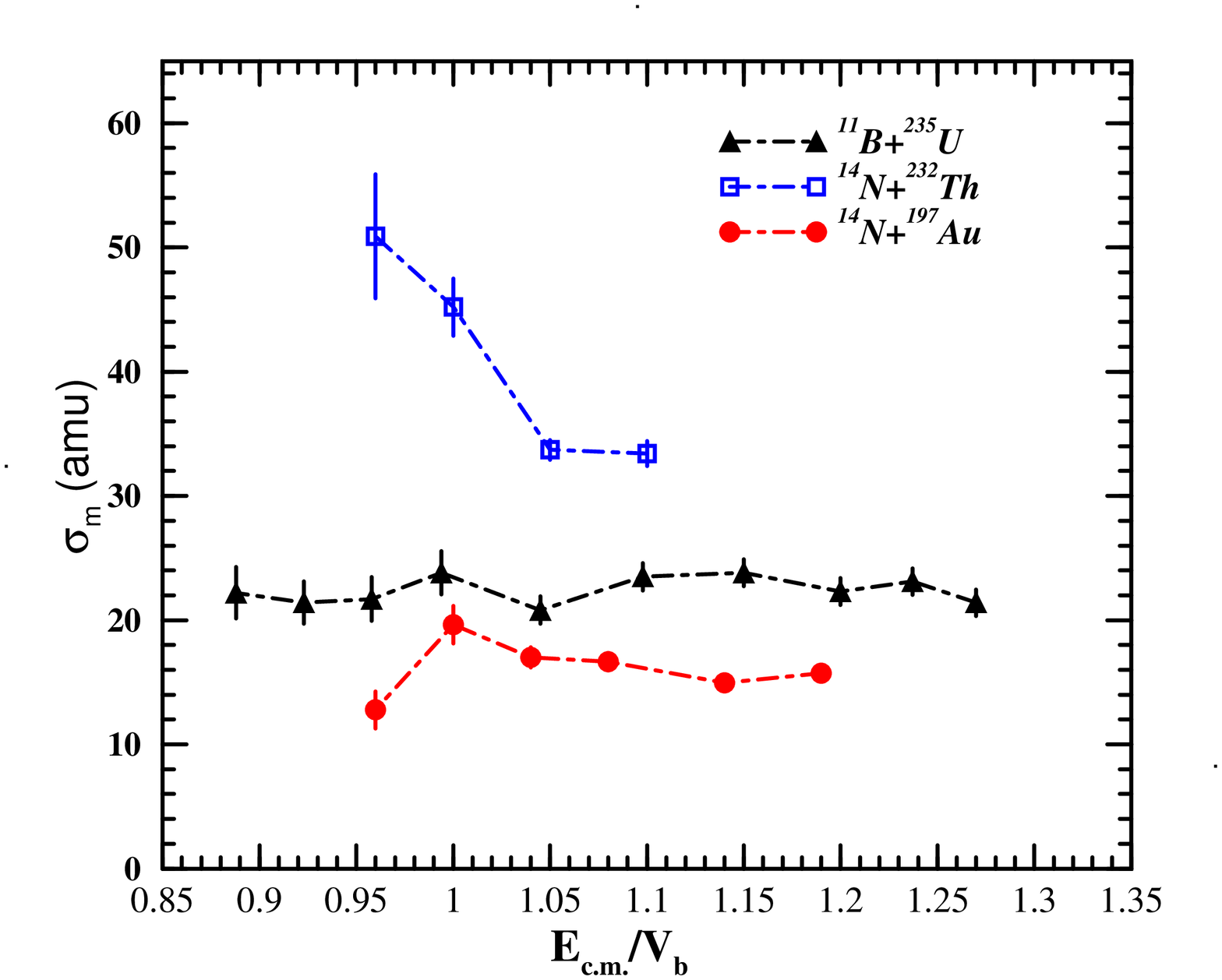}
\caption{\label{fig:fig3}~ (Color online) Variation of the standard deviation  $\sigma_{m}$ to the fitted Gaussian of the fission fragments mass distribution as a function  of  $E_{\rm c.m.}/V_b$ }
\end{figure}

\begin{figure}
\includegraphics*[scale=0.75,angle=0]{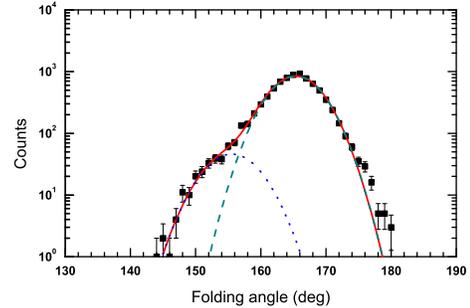}
\caption{\label{fig:fig4}~ (Color online) Measured fragment folding angle distribution for the system $^{14}$N+$^{232}$Th at projectile energy $E_{\rm c.m.}$= 73.5 MeV }
\end{figure}

The variation of standard deviation of the mass distributions with excitation energy for two reactions  $^{14}$N + $^{232}$Th and $^{11}$B+$^{235}$U forming the same composite system $^{246}$Bk is shown in Fig.~\ref{fig:fig5}. Solid (red) line in the figure shows the calculated variation from statistical theory \cite{Back96}  following the relation $\sigma_m^2$ = $\frac{1}{k}$ {$\sqrt{\frac{E^{\dag}}{a}}$, where $E^{\dag}$ is the excitation energy at the scission point, $a$ is the nuclear level density parameter. A value of the stiffness parameter k =0.0033 MeV/$u^2$ fitted the $^{11}$B+$^{235}$U data well \cite{Itkis}. It is interesting to note that  for the $^{14}$N + $^{232}$Th reactions, not only there is a sudden jump of mass widths ($\sigma_m$) near the Coulomb barrier but also the magnitude is higher than the $^{11}$B+$^{235}$U reactions over the entire range of excitation energies.

\begin{figure}
\includegraphics*[scale=0.35,angle=-90]{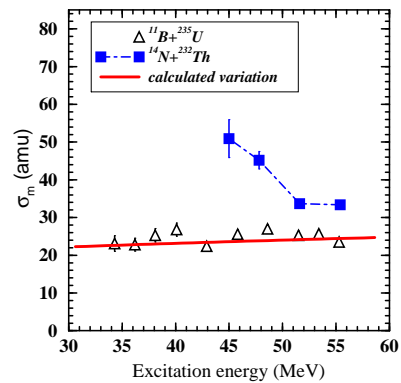}
\caption{\label{fig:fig5}~ (Color online) Measured variation of $\sigma_{m}$ with excitation energy  for the two reaction forming the same composite system $^{246}Bk$. Calculated variation is shown by solid (red) line.}
\end{figure}

The width of fission fragment mass distribution mainly depends on the excitation energy. However, it has weaker linear dependence on the mean square angular momentum $<l^2>$   brought in by the projectile \cite{Dubna07}. In Fig.~\ref{fig:fig6} we show the measured \cite{BeheraRapid} variation of $<l^2>$  with excitation energy for the two systems forming the same composite system $^{246}$Bk . It  can be seen that  average mean square angular momentum value for the system $^{11}$B+$^{235}$U is always higher than those for  $^{14}$N + $^{232}$Th in the range of our measured excitation energies. So, even if we correct the predicted widths of the mass distribution for the possible contribution for $<l^2>$, the correction for the $^{14}$N + $^{232}$Th system would be smaller compared to that for $^{11}$B+$^{235}$U and would not explain the observed  sharp increase of the mass width for  $^{14}$N + $^{232}$Th around the Coulomb barrier as shown in Fig.~\ref{fig:fig5}.

\begin{figure}
\includegraphics*[scale=0.4,angle=0]{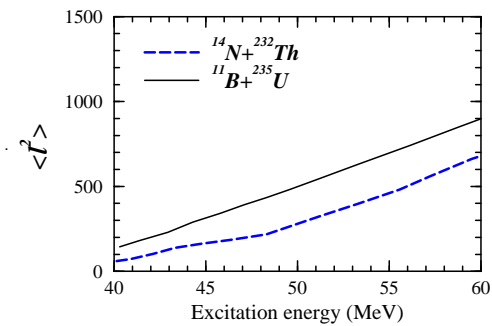}
\caption{\label{fig:fig6}~ (Color online) Variation of mean square angular momentum with excitation energy  for the nucleus $^{246}Bk$, populated by the two different entrance channels.}
\end{figure}

We have observed that for the fusion of two heavy nuclei forming the compound nucleus $^{246}$Bk through the reactions  $^{11}$B+$^{235}$U and $^{14}$N + $^{232}$Th at the similar excitation energy  the fused system behaves differently and the reactions paths followed by the two systems, particularly around the Coulomb barrier, are significantly different. The targets in both cases are prolate deformed in the ground state while the projectiles are spherical; yet, the reaction paths followed by the two systems are quite different.  In the case of $^{11}$B, the composite $^{246}$Bk shows a behaviour according to the predicted course following statistical equilibration of the composite to a compound nucleus followed by binary fission. This reaction mechanism holds true at near and below the Coulomb barrier and the variation of the width of the mass distribution closely follows that predicted from the statistical laws as shown by the solid  (red) line in Fig.~\ref{fig:fig5}. On the contrary, this expected reaction mechanism does not hold true for the  $^{14}$N + $^{232}$Th system, as the mass widths at all the measured energies are larger than those predicted by the statistical theory and show a sharply increasing trend as the energy is lowered than about 1.05 times the Coulomb barrier.

We have earlier reported anomalous behaviour of the mass width, increasing with the decrease in energy, for the reacting systems of $^{12}$C, $^{16}$O and $^{19}$F on $^{232}$Th \cite{myPLB}. Although these three reactions did not produce the same composite system, we could explain the increase in the mass widths in terms of a orientation dependent quasi-fission reaction mechanism. For the $^{14}$N + $^{232}$Th system, in the measured energy region, the reactions proceed mostly through the impact of the projectile on the polar region of the target nucleus and it drives the non-compact entrance channel shape to an almost mass symmetric saddle shape leading to increased mass widths, rather than a compact equilibrated fused system, which undergoes fission through shape oscillations. We can explain the observed increase of the mass widths with decrease in energy, in the case of  $^{14}$N + $^{232}$Th system, as a quasi-fission phenomenon, however, in case of the $^{11}$B+$^{235}$U system, the initial separation in the entrance channel may not be sufficient to drive the system to a non-compact entrance channel to a mass asymmetric saddle shape. We also point out that, although the entrance channel mass asymmetries for the two systems do not differ much, in the two cases the flow of mass is completely different as the two nuclei fuse together. In the system $^{14}$N + $^{232}$Th,  the entry point to the multidimensional potential surface is such that the system is driven to lower mass asymmetry leading to greater probability to diverge to a mass asymmetric saddle rather than a compact fused system, while the picture is just the reverse in the case of $^{11}$B+$^{235}$U, whereby the system is driven to higher mass asymmetry and thus is more probable to reach a compact fused system and undergo statistical fission after equilibration.

We conclude that in the fusion of the $^{246}$Bk, in addition to the effect of deformation, where the reaction proceeds mostly through impact through the polar region of the prolate thorium or uranium targets, the entrance channel mass asymmetry plays a crucial role in the reaction mechanism, particularly in the energy close to the Coulomb barrier. Even a small change in the entrance channel mass asymmetry, which effectively reverses the flow of mass in fusing the target and projectile, results in completely different reaction mechanism - a re-separation of the composite in $^{14}$N + $^{232}$Th over a mass asymmetric saddle, or the statistical equilibration followed by fission in the case of $^{11}$B+$^{235}$U. The entrance channel sharply affects the fusion process in the production of the nucleus $^{246}$Bk .

We are thankful to Dr. A. Saxena of BARC, Mumbai and Dr S. K. Das of VECC for providing help of making few of the targets used in the experiment. Thanks are due to the staff members of the IUAC Pelletron for providing good quality pulsed beam required for the experiment.

\newpage


\begin{thebibliography}{99}
\bibitem{Ognessian} Yu. Ts. Oganessian {\sl et al.}, Phys. Rev. C {\bf 74} 044602 (2006).
\bibitem{ItkisNPA} M. G. Itkis {\sl et al.}, Nucl. Phys. A {\bf 734} 136 (2004). 
\bibitem{KailashPR} S. Kailas , Phys. Rep. {\bf 284} 381 (1997).
\bibitem{RamPRL85} V. S. Ramamurthy and S. S. Kapoor, Phys. Rev. Lett. {\bf 54}, 178 (1985).
\bibitem{myPLB} T. K. Ghosh {\sl et al.}, Phys. Lett. B {\bf 627} 26 (2005).
\bibitem{Renju08} R. G. Thomas {\sl et al.}, Phys. Rev. C {\bf 77} 034610 (2008).
\bibitem{MollerNature} P. Moller and A. J. Sierk, Nature {\bf 422} 485 (2003). 
\bibitem{HindePRL95} D. J. Hinde {\sl et al.}, Phys. Rev. Lett. {\bf 74} 1295 (1995).
\bibitem{Rafiei} R. Rafiei {\sl et al.}, Phys. Rev. C {\bf 77} 024606 (2008).
\bibitem{BeheraRapid} B. R. Behera {\sl et al.}, Phys. Rev. C {\bf 64} 041602(R) (2001).
\bibitem{Singh08} BirBikram Singh {\sl et al.}, Phys. Rev. C {\bf 77} 054613 (2008).
\bibitem{Businaro} U. L. Businaro and S. Gallone, Nuovo Cimento {\bf 5}, 315 (1957).
\bibitem{myRapid2} T. K. Ghosh {\sl et al.}, Phys. Rev. C {\bf 70} 011604(R) (2004).
\bibitem{myNIM} T. K. Ghosh {\sl et al.}, Nucl. Instrum. Methods Phys. Res. A {\bf 540} 285 (2005).
\bibitem{Back96} B. B. Back {\sl et al.}, Phys. Rev. C {\bf 53} 1734 (1996).
\bibitem{Itkis} M. G. Itkis {\sl et al.}, Yad. Fiz. {\bf 52} 23 (1990).
\bibitem{Dubna07} G. N. Knyazheva {\sl et al.}, Phys. Rev. C {\bf 75} 064602 (2008).

\end{thebibliography}
\end{document}